# High-efficiency, 80-mm aperture metalens telescope

Lidan Zhang,[1,†] Shengyuan Chang,[1,†] Xi Chen,[1] Yimin Ding,[1] Md Tarek Rahman,[1] Yao Duan,[1] Mark Stephen,[2] and Xingjie Ni[1,*]

[1] *Department of Electrical Engineering, the Pennsylvania State University, University Park, PA 16802*

[2] *NASA-Goddard Space Flight Center, Greenbelt, MD 20771*

[*] xingjie@psu.edu

[†] These authors contributed equally to this work.

**Abstract**

Metalenses, artificially engineered subwavelength nanostructures to focus light within ultrathin thickness, promise potential for a paradigm shift of conventional optical devices. However, the aperture sizes of metalenses are usually bound within hundreds of micrometers by the commonly-used scanning-based fabrication methods, limiting their usage on practical optical devices like telescopes. Here, for the first time, we demonstrate a high-efficiency, single-lens, refractive metalens telescope. We developed a mass production-friendly workflow for fabricating wafer-scale (80-mm aperture) metalenses using deep-ultraviolet (DUV) photolithography and a multi-exposure process involving reticle rotation and pattern stitching to leverage the radial symmetry of metalenses. Our metalens works in the near-infrared region (1200 – 1600 nm) with diffraction-limited performance and a high peak focusing efficiency of 80.84% at 1450 nm experimentally. Based on the metalens, we built a single-lens telescope and acquired images of the lunar surface, revealing its geographical structures. We believe our demonstration of the metalens telescope proves the exciting potential lying in the metasurfaces and could bring new possibilities for areas involving large optical systems, including geosciences, planetary observation, and astrophysical science.

**Introduction**

Not long after its first appearance, the telescope found its place in astronomy[1] and became our sail to reach for the stars ever since. To this day, humans have constructed telescopic systems that are both land-based and spaceborne, revealing valuable insights into our galaxy and the universe. However, the resolving power of telescopes heavily relies on their aperture size, leading to the extreme weight and footprint that bring challenges to their construction and deployment. Similar problems also occur for conventional optical systems relying on complicated multi-element designs to achieve the desired level of optical performance, where elements made of glasses or other refractive materials are bulky and heavy. Moreover, lens groups consisting of two or more lenses are often cascaded together for aberration correction, leading to a precipitous increase in the size and weight of the system [2–4]. The solution to these problems may surf with recent advances in nanofabrication techniques and the so-called "metasurfaces" – artificial interfaces that mold light with spatially varying subwavelength nanoantennas, *i.e.*, meta-atoms. Being ultrathin and lightweight, metasurfaces have provided a new approach to recasting optical components into flat devices without performance deterioration. [5–8] In particular, metasurface lenses, *a.k.a.* metalenses, have attracted tremendous attention in the past decade for their potential to surpass conventional lenses with ultra-thin planar configurations, which promise a drastic reduction in optical systems' size and weight. [9–11] Moreover, metalenses can realize optical functionalities beyond conventional bulky lenses due to localized light manipulation with subwavelength precision. In addition, metalenses could realize dispersion engineering from both chromatic and angular perspectives, which can be utilized to design optical devices ranging from chromatic-aberration-corrected singlets or dispersion-enhanced metalenses for hyperspectral imaging. [12–16] However, despite numerous advantages mentioned above, the application of metalenses, or in general meta-optical devices, remains elusive due to their limited aperture sizes.

A major reason for this limit lies in the manufacturing processes of meta-optical devices. Currently, electron-beam (e-beam) lithography and focused ion beam (FIB) are commonly used to define the nanoscale, subwavelength meta-atoms thanks to their power to resolve nanometer-scale features with high precision. However, being scanning-based techniques, e-beam lithography and FIB come with slow speed, high expense, and limited scalability, making it challenging to fabricate large-aperture meta-optical devices even for mere prototyping, let alone mass production. [17,18] To date, most of the metalenses reported have only (sub-) millimeter scale apertures, while conventional lenses with at least centimeter-scale apertures are commonly seen in our daily life, even considered entry-level for applications like telescopic imaging. This also implies that if the metasurface can be made with a large aperture and high efficiency, it will become a game-changer in areas previously considered inaccessible to meta-optics, such as planetary observation and remote sensing and imaging. [19–25] Motivated by this, we addressed this challenge by developing a wafer-sized high-efficiency metalens to build a telescope, along with a complete fabrication workflow to enable its manufacture in a cost-effective, mass-producible fashion.

Over the past years, a few attempts have been made to develop cost-effective and scalable fabrication techniques for manufacturing large-aperture metalenses. For example, the nano-imprint method has been used to fabricate gallium nitride metalenses with an aperture size of ~1 cm[26]. Although the nano-imprint technology is mass-production friendly, it suffers from pattern defects, low throughput, and template wear[27]. An alternative method for large-scale metalens fabrication is photolithography, which is known for its superior throughput. Using stepper lithography, metalenses with aperture sizes up to 2 cm have been demonstrated in near-infrared and visible wavelengths[28–30]. However, those metalenses are still way smaller than their conventional counterparts. In addition, while most photolithography methods are considered much more efficient in fabricating large-area patterns, their resolutions are limited by the wavelength of the light source, which is much larger than that of the electron beam. They may fail to resolve the fine features of meta-atoms and deteriorate their performance.

Here, for the first time, we demonstrate an 80-mm aperture, high-efficiency, refractive metalens telescope capable of observing celestial bodies such as the moon (Figure 1). The metalens, the core part of our telescope, was first designed numerically and then fabricated on a four-inch silicon wafer with deep-ultraviolet (DUV) projection stepper lithography (top right inset, Figure 1), which is commonly used in the semiconductor industry. To tackle the aforementioned problems of photolithography, we combined multiple exposures with pattern stitching and rotation. The fabrication was conducted with projection stepper lithography using a 248-nm DUV light source. The translation and rotation of the wafer were enabled by a mechanical step-and-repeat wafer stage with laser interferometry and a computer-controlled stage motion system. Our efforts resulted in a single transmissive metalens with an unprecedented 80-mm aperture (bottom left inset, Figure 1).

Following the fabrication, the performance specifications of the metalens were characterized to determine whether it could meet our need for a telescope objective. Its focusing and imaging performance was tested through a series of experiments conducted both indoors and outdoors. The focusing measurement results indicate that our metalens achieved a diffraction-limited performance (Strehl ratio ~85%) ranging from 1200 nm to 1600 nm with varying focal lengths since the metalens was designed for single-wavelength operation. The focusing efficiency of our metalens reached a peak of 80.84%, measured at the wavelength of 1450 nm. Further imaging and video recording tests with either a standard resolution test target or real-life objects demonstrated superior resolving power and light collection efficiency of our metalens. With these results, we finally went ahead to build the metalens telescope and acquired images of the moon's surface. To the best of our knowledge, this is the very first-time metalens were used for celestial imagery.

We think that our demonstration could raise the attention to applications of metasurfaces on a broader horizon, and we expect our work to reveal a viable way for mass-production of large-scale meta-optic devices. We believe the result of our work could benefit metalenses and their applications in various areas, including display technologies, remote sensing systems with satellites, and imaging systems for airborne vehicles.

**Results**

Our large-aperture metalens, the core of the metalens telescope, was designed by calculating the required phase profile $\varphi(r) = -\frac{2\pi}{\lambda}[\sqrt{(r^2 + f^2)} - f]$, where $f$ is the focal length of the metalens, $\lambda$ is the wavelength of the incident light, and $r$ is the radial position on the lens. We designed nanocylinder antennas, which are the meta-atoms of our metalenses, made of amorphous silicon on a fused silica substrate. Simulating the nanocylinders using a full-wave solver, we found that a complete $2\pi$ phase coverage can be achieved with nanocylinders' diameters ranging from 240 to 520 nm (Figure 2(a)). A resonance near 400 nm is excluded to ensure a smooth change of phase. In addition, only antennas with greater than 95% transmittance were chosen to ensure the high efficiency of the resulting metalenses. A small metalens with a diameter of 40 μm was designed and simulated, where the simulation results match well with the theoretical calculation, validating our meta-atom designs (see supplemental material).

In contrast to the e-beam lithography, there is a large discrepancy between the sizes of designed features and the fabricated ones in our photolithography approach as our metalens' minimum feature size is determined by the resolution limit of both mask writing and photolithography processes. In order to achieve accurate sizes for our pattens, we first established the relationship between the input meta-atom dimensions and the fabricated meta-atom dimensions measured in scanning electron micrographs (SEMs) by multiple dose and etching tests (Figure 2(b)). We then use a quadratic fitting to relate the meta-atom dimensions in the layout and the fabricated patterns. The meta-atom dimensions in the final layout were corrected following the established quadratic relation, which guarantees that the fabricated antennas' dimensions match the required ones. Then, compression of the layout data was carried out since a metalens of 80-mm-diameter aperture easily contains over one billion meta-atoms, making the layout file size excessively large for any practical read/write actions. In order to realize the compression, we take advantage of the radial symmetry of the metalens phase profile and store a lookup table (LUT) with the positions and corresponding meta-atom designs of which the circular cross-sections were approximated with polygons. The designs of the meta-atoms were determined based on their positions via the LUT. The dimension of the nanocylinder in the LUT follows a specific increment, which can be adjusted to control the precision of the lookup process. For our 80-mm metalens, the layout file is about 240 GB with a 5-nm increment. Our pattern layout files were generated corresponding to the required phase profile with both correction and compression.

We fabricated our metalens with our customized DUV lithography workflow. The fabrication process of our large-aperture metalens included two parts – reticle fabrication and metalens fabrication. Considering the maximum area for a single exposure of the DUV tool is a 22-mm-by-22-mm field, we divided our patterns into 16 pieces, with each piece a 20 mm × 20 mm square for exposure. Exploiting the $C^4$ rotational symmetry of the metalens, only four reticles are needed to cover a quarter of the metalens, simplifying the lithography process. The four reticles were made with the pattern size magnified four times for the projection lithography (Figure 3(a), (d)). Our metalens pattern was transferred onto a four-inch fused silica wafer by a DUV stepper with four reticles at a 4:1 reduction ratio (Figure 3(a)). The whole exposure process for 16 parts can be done either by reticle rotation (rotation markers needed on the reticles) or wafer rotation (rotation markers needed on wafers) with appropriate stitching for different fields. In the exposure process, four reticles were loaded into the Stepper. A quarter of our metalens pattern, consisting of four 20 mm × 20 mm fields, was exposed through four different reticles. The other three quarters were exposed by rotating the relative position of reticles and wafers by 90°, 180°, and 270° (Figure 3(b)). The exposed pattern was followed by post-exposure bake and development and then transferred to the amorphous silicon layer below through the inductively coupled plasma reactive ion etching process. The entire process flow is shown in Figure 3(c) and the fabricated metalens is shown in Figure 3(e). We observed that the stitching error among different fields was less than 1 μm (see supplementary materials), which has negligible influence on the performance of the metalens. Characterization of the fabricated metalens by SEM shows that the constituent nanocylinder antennas match well with our design (Figure 3(f) and top left inset of Figure 1). In summary, our DUV photolithography-based fabrication method embraces both precision and simplicity while eliminating the need for scanning-based pattern writing, thus significantly reducing cost and time and being suitable for mass production commercial telescopes.

With the metalens fabricated, we characterized the optical performance specifications of the fabricated metalens with a homemade optical platform. For focusing measurements, the focal spot image was captured using an imaging system, as shown in Figure 4(a) (see Methods section). By tuning the output wavelength of the incident light and translating the objective lens along the optical axis, we characterized the intensity profile of the focused beams along both yOz and xOy cross-sections over a broad wavelength range, from 1200 nm to 1600 nm (Figure 4(b)). Our measurement shows that, although the metalens was designed to operate at the 1450 nm wavelength, it could still focus light tightly across the tested wavelength range, with varying focal lengths due to material dispersion. The measured focal lengths show a roughly linear dependence with the incident wavelength (Figure 4(c)). Further, the quality of each focal spot was evaluated by extracting its full width at half maximum (FWHM) and comparing it to that of a diffraction-limited system with the same numerical aperture (NA). The results show that the measured FWHMs were close to diffraction-limited values across all wavelengths (Figure 4(d)). Note that the NAs used for determining the diffraction-limited spot sizes were calculated based on the metalens aperture size and the measured focal lengths at each wavelength. In addition, the Strehl ratio of our metalens' point spread function (PSF) reaches ~ 0.85 at all wavelengths, indicating the diffraction-limited performance of our metalens (It is widely accepted that an optical system with a Strehl ratio above 0.8 is considered to have diffraction-limited performance). Finally, we measured the focusing efficiencies of our metalenses using an optical power meter. Since the metalens aperture greatly exceeds the active area of the power meter photodetector, and the focal spot size of the metalens was far smaller than the same active area size, we mounted a pinhole in front of the photodetector. The purpose of the pinhole is twofold: (1) to sample the intensity distribution in front of the metalens and fit with a two-dimensional Gaussian profile to determine the incident light power; (2) to limit the detector aperture to only capture light within an area across about three times of the FWHM of the focal spot size at the metalens focal plane. Details of the efficiency measurement can be found in the supplementary material. Our measured focusing efficiency reaches its maximum at 1450 nm, with a value of 80.84% (Figure 4(f)). Compared to previously reported works, this efficiency is among the highest values reached by metalenses operating at a single wavelength. These results suggest that, with a diffraction-limited PSF as well as an excellent efficiency, our metalens can serve as a telescope objective.

Next, we tested the imaging performance of our metalens in both laboratory and outdoor environments to determine its resolving power. First, we imaged the USAF 1951 resolution test chart with the single-lens imaging configuration, and the as-captured grayscale image (Figure 5(a-b)). A beam expander and bandpass filter were used to ensure uniform monochromatic illumination (see Methods section). The smallest feature we could resolve is Group 5, Element 6 in the resolution test chart, with a line width of 8.77 μm (57 line pairs/mm). This value is close to the focal spot FWHM we measured earlier, which means we had reached the limit of the resolving power of our metalens. Second, we used the same configuration to image a burning candle. We captured images (Figure 5(c)) and recorded videos of the candle frame (see Supplementary Movie S1).

Finally, we built our metalens telescope with the fabricated metalens as the objective. The image was magnified 2x by a 4-f system before being captured by the camera to fully utilize its pixels. The system was assembled on an optical breadboard mounted on a commercial camera tripod. Through this metalens telescope, we imaged the lunar surface in an outdoor environment (Figure 5(d-e)) with the maximum frame rate (60 fps) of our camera. We identified some features of the lunar surface, such as the Sea of Rains, the Sea of Serenity, the Sea of Tranquility, the Ocean of Storms, the Sea of Tranquility, and the Copernicus crater, from the obtained image. Note that the blurry right edge of the moon was attributed to the Waning Gibbous phase of the moon. We could determine that our metalens telescope's minimum resolvable feature size is ~ 80 km on the lunar surface, which has *never* been achieved by *any* previously reported metalenses. Thanks to the large aperture of our metalens, we demonstrated a possible application of metalenses in astronomical observations, which further levitates the potential of metasurface technology.

**Discussion**

In conclusion, we developed a metalens telescope system capable of celestial imagery based on a high-efficiency, 80-mm aperture metalens. In order to fabricate such large metalens, we leveraged the radial symmetry of the pattern and proposed an approach to fabricate high-performance, large-aperture metalenses with reduced expense and time, which involves the rotation-and-stitching of the pattern and multiple-exposure processes with DUV photolithography. The application of DUV photolithography allows us to skip the costly, time-consuming scanning-based lithography processes for large-area patterns with densely packed nanostructures, and the symmetries of the metalens pattern are exploited to significantly reduce the layout file size as well as the number of reticles needed for photolithography. This developed process can be readily extended for the mass production of large-scale meta-optical components. Through our characterization, we observe that our wafer-scale (80-mm diameter) metalenses have diffraction-limited performance and over 80% peak focusing efficiency at 1450 nm. Due to chromatic aberrations, shifting of focal length was observed at other frequencies, but focal spots formed had Strehl ratios ~ 85% across the near-infrared region from 1200 to 1600 nm, indicating diffraction-limited performance. Finally, the metalens undergoes a series of imaging tests to determine its resolving power, and are used in the metalens telescope to acquire images of the lunar surface.

Metalens is inherently spherical aberration free and its ultrathin thickness makes them ideal as an alternative to the bulky lens groups, which are an essential part of modern telescope systems for aberrations correction. Although our metalens telescope is limited to single wavelength imaging in near infrared, the metalens telescope can be further improved with achromatic designs for better imaging performance. Besides, the resolution of DUV lithography can be further improved by reducing the feature size in reticle fabrication by advanced mask writer and carefully control the DUV conditions, which provides possibility for extending the working wavelength to visible light. Furthermore, DUV lithography allows precise alignment from both the front and the back side, which provides the possibility for multilayer stacking of metasurfaces with different functionalities for higher level of integration.

With the growth of market demand for compact optical systems, the need for lightweight, cost-effective, and high-quality large-aperture optical components is dramatically increasing. Our work forecasts the paradigm shift brought by lightweight large-aperture metalenses and other meta-optical devices. The demonstrated metalens telescope could be used as a compact, versatile, and energy-efficient imagery sub-system for remote sensing systems or infrared imaging systems for aircraft, unmanned vehicles, drones, and satellites. The developed large-scale metalenses can be readily integrated into existing optical systems for reduced footprints and weights, and can also be used in consumer products such as virtual reality (VR), augmented reality (AR), and mixed reality (MR) devices. The developed fabrication technique could lead to highly scalable, cost-effective mass production of large-aperture metalenses and can be readily extended to other large-scale patterns of optical nanostructures. We believe our work could potentially impact the greater metamaterials, electromagnetics, and optical research communities.

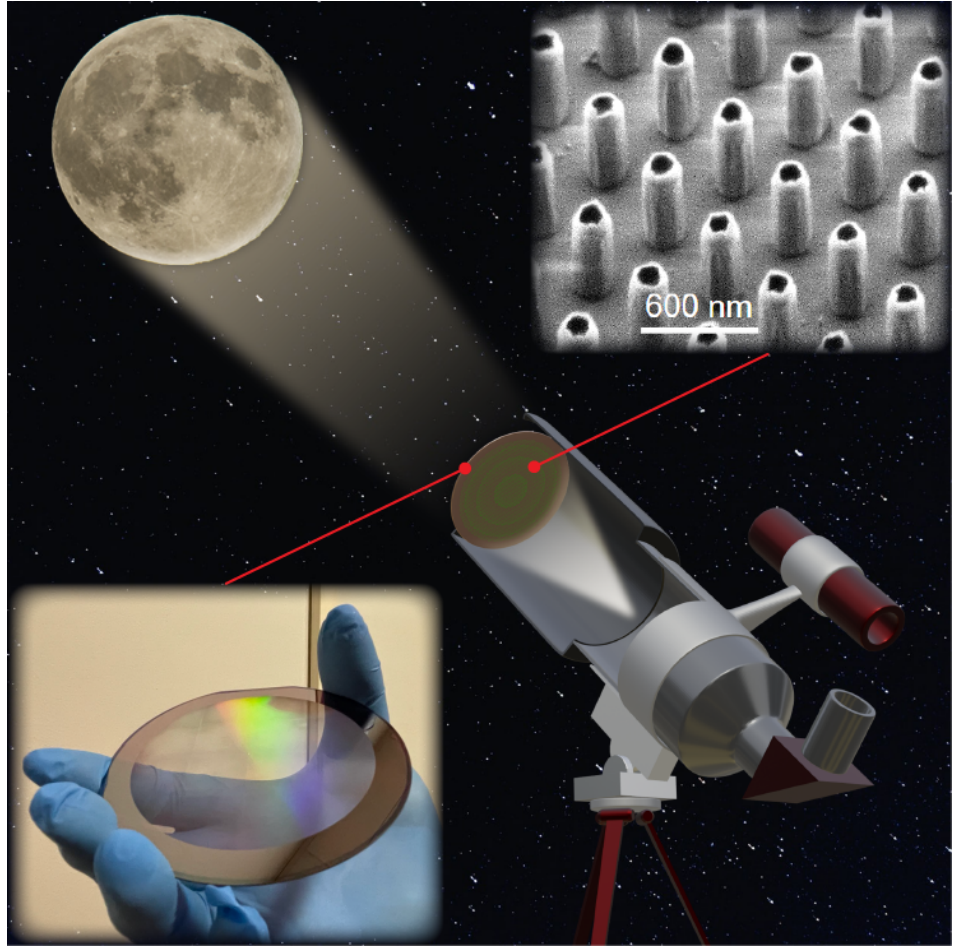

**Figure 1. The metalens telescope, with an 80-mm metalens as its objective.** A schematic illustration of the metalens telescope. The metalens is used as the objective of the telescope and focuses light from the moon to form its image. The telescope tube is shown only for illustrative purposes and does not exist in our real setup. The image capture system, including a camera and a filter, is also omitted for simplicity. Background picture: from Paul Volkmer via Unsplash. Image of the moon: from NASA's Solar System Exploration website (solarsystem.nasa.gov). (Top inset) An SEM image of its constituent meta-atoms, scale bar: 600nm. (Bottom inset) A photograph of the fabricated wafer-scale metalens.

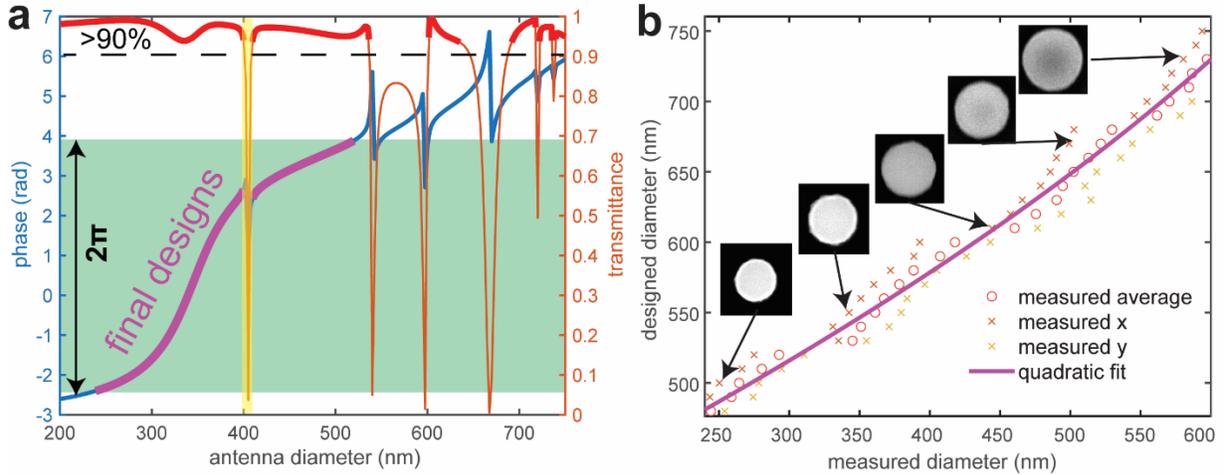

**Figure 2. Design principles and size-correction method for the meta-atoms.** (a) The simulated phase and transmittance response of amorphous silicon nanocylinders with different diameters ranging from 200 nm to 800 nm. The antennas with diameters ranging from 240 nm to 520 nm were selected as the final designs (purple line inside green shaded area), covering a complete $2\pi$ phase range. A resonance regime around 400 nm diameter was excluded due to the low transmittance and rapid phase changes. The transmittance for our designs is over ~95%. (b) fabrication test result for the relationship between the measured diameter of the fabricated nanocylinders and the designed diameters in the input layout. With the appropriate dose for the exposure, our smallest design (240 nm diameter) can be fabricated with a pattern size of 480 nm, corresponding to 1.9 µm in the reticles (magnified by 4 times), which the mask writer can achieve. (Insets) SEM images of the fabricated nanocylinders of different sizes.

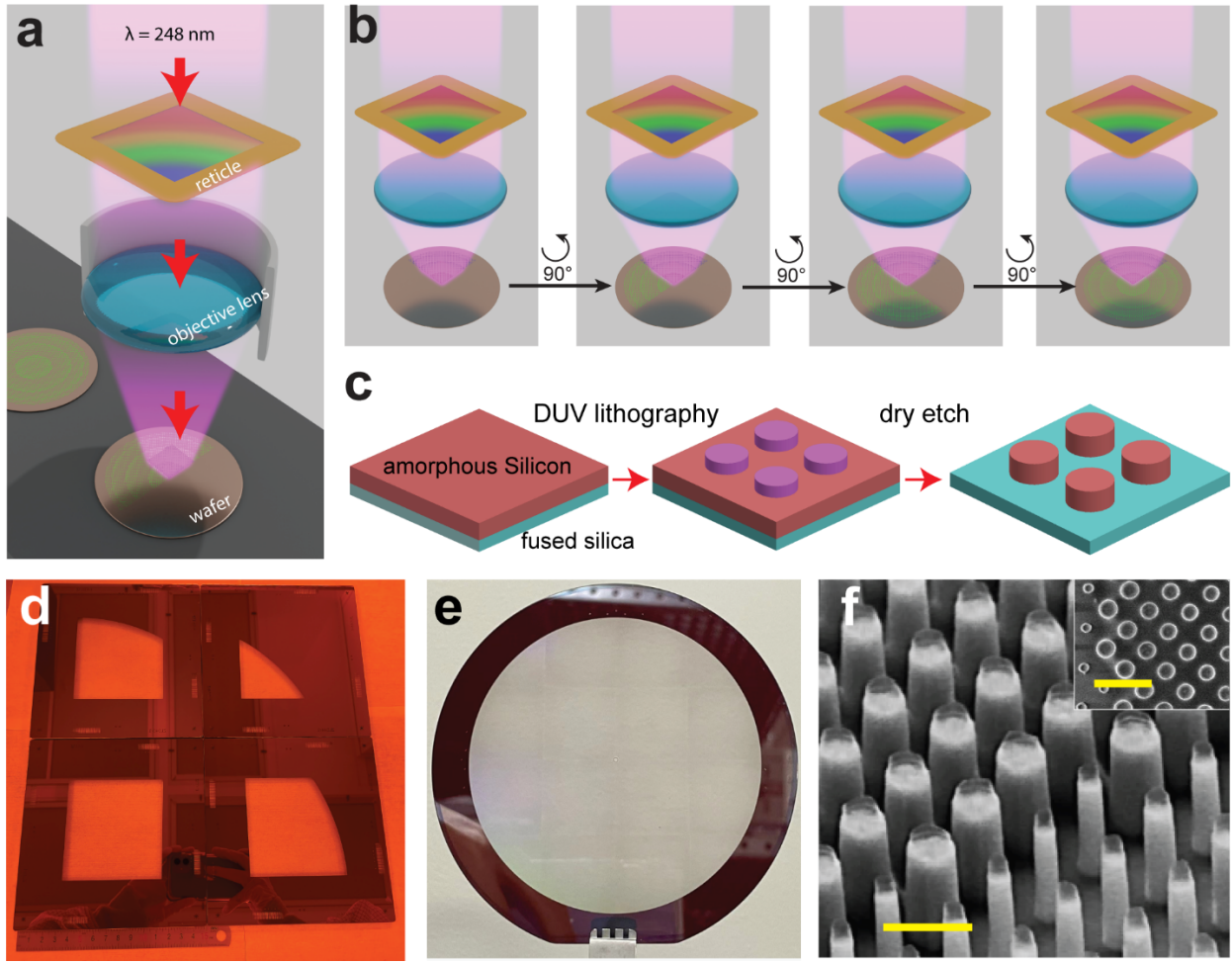

**Figure 3. Fabrication results of the wafer-scale metalenses.** (a) A schematic illustration of the exposure process of DUV lithography for large-scale metalenses fabrication. A light source with a 248 nm wavelength is used for the DUV stepper system. The patterns on the reticles (photomasks) are projected to the wafer through projection lenses with 4:1 reduction, generating up to a 22-mm-by-22-mm area with a single exposure. The projection process is fast, and hundreds of wafers can be made per hour, suitable for mass production. (b) The rotating method is used during DUV exposure. By rotating the relative position of the reticle and the wafer with prefabricated rotation markers by 90°, the exposure can be realized in different quadrants of the metalens. Either reticles or wafers can be rotated. But the reticles can only rotate for 90° each time in the loading process, while the wafers can be rotated for an arbitrary angle using the stage controller, which provides more flexibility. (c) A schematic of the fabrication process flow. A thin amorphous silicon film was deposited on a fused silica wafer. DUV lithography was followed to make the resist pattern, and then the resist pattern was transferred to amorphous silicon film by dry etching. (d) A photograph of the four reticles used in the DUV lithography process. (e) A photograph of the final 80-mm-aperture metalens. (f) A tilted-beam scanning electron micrograph (SEM) of metalens, revealing nanocylinders of different sizes and vertical sidewalls. (Inset) A top-view SEM of the metalens. Scale bars: 1 μm.

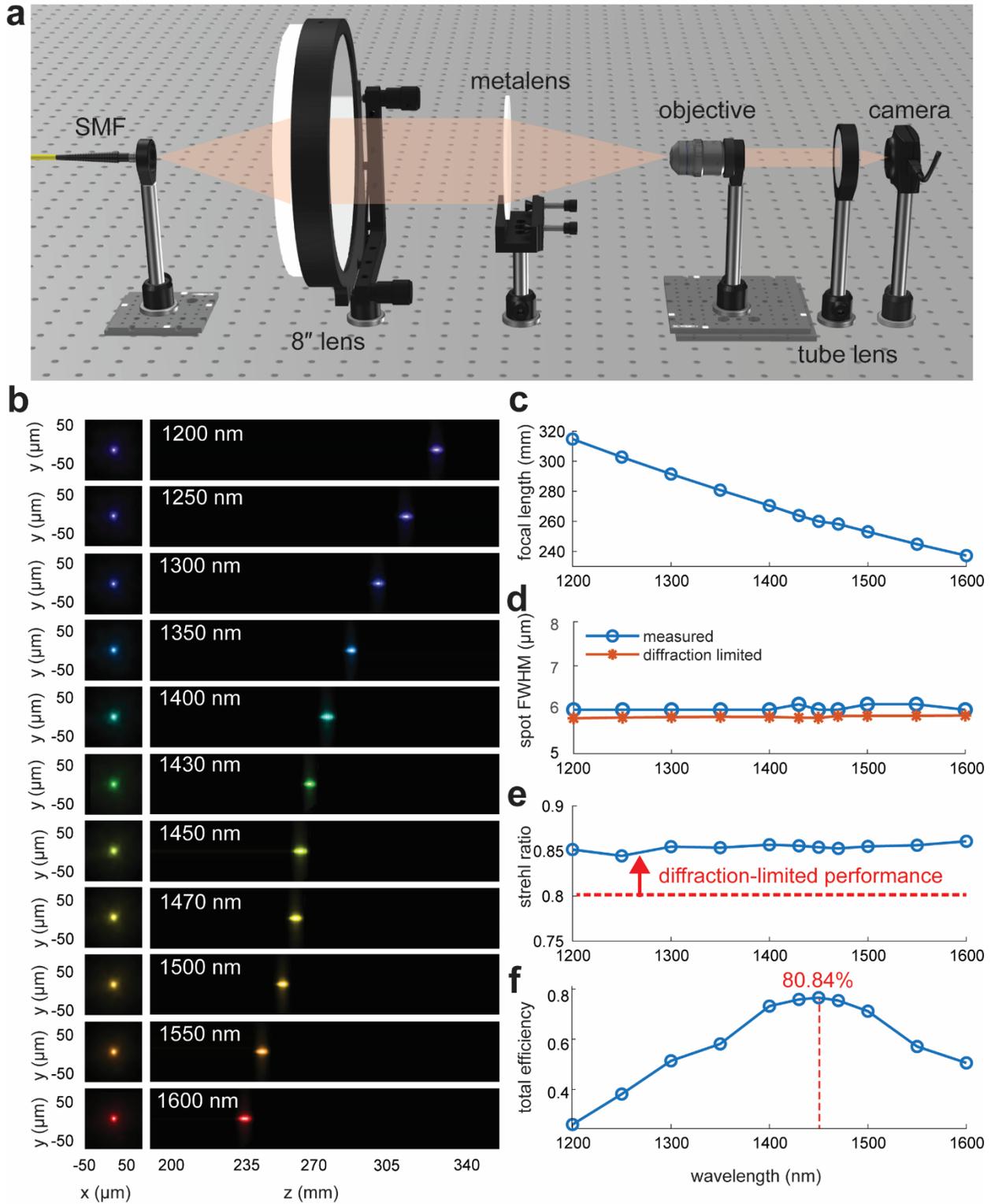

**Figure 4. Focusing performance of the fabricated metalens.** (a) A schematic of the experimental setup for characterizing the focusing performance of the metalens. The laser out-
15

coupled from a single-mode fiber was expanded and collimated by an 8-inch lens to ensure the incident beam size is much larger than the metalens aperture. (b) Intensity profiles of the focused beam after the metalens along both xOy and yOz cross-sections with the wavelength ranging from 1200 nm to 1600 nm. (c-f) Measured focusing properties of the fabricated metalens with the wavelength ranging from 1200nm to 1600nm: (c) The focal lengths, (d) focal spot FWHM, (e) point spread function Strehl ratio, and (f) total focusing efficiency.

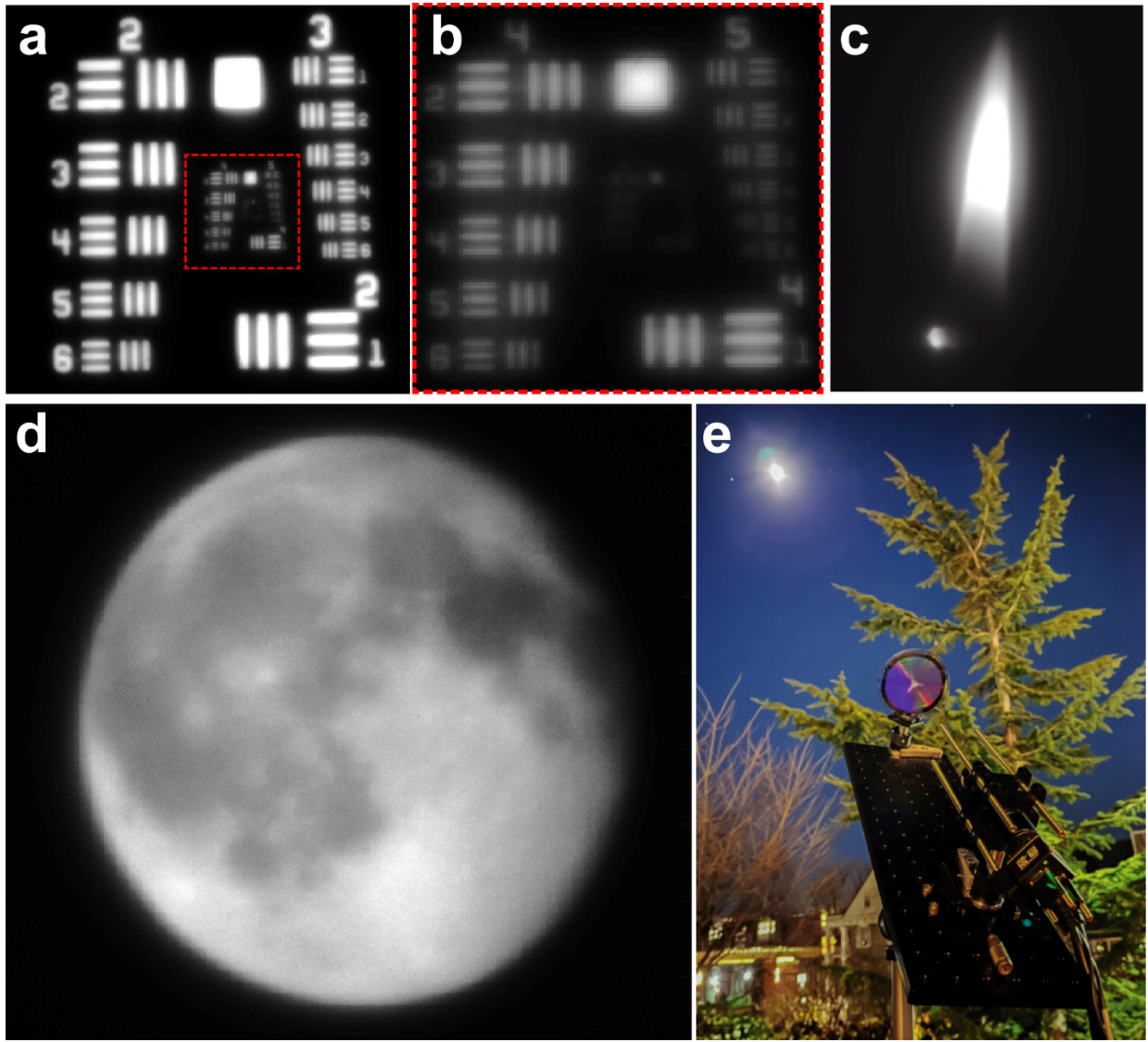

**Figure 5. Imaging performance of the fabricated metalens.** (a) Captured images of the USAF 1951 resolution chart formed by the 4-inch-aperture metalens. (b) Zoomed-in view of the red boxed area in (a), showing the center area consisting of patterns from Group 4 and above. (c) Infrared image of the flame from a lit candle. (d) An IR image of the moon formed by the metalens. The image was magnified 2x by a 4-f system in front of the camera. (e) A photograph of the metalens telescope which was being used to capture the moon's image. The same bandpass filter was used to reduce the background noise. All images captured by the camera are without any post-processing.

## Materials and Methods

### Reticle fabrication.

We used a 6" × 6" × 0.250" photomask blank (Telic company) with a low reflective chrome layer (8% reflective @ 450nm) and 530-nm AZ 1500 photoresists for the fabrication of reticles. After writing the metalens patterns (magnified by 4×) and the alignment markers for rotation on photomasks with a DWL2000 laser writer, AZ 300 MIF developer was used to develop the photoresist layer. The final Cr patterns were wet etched with the resist masks, and the reticles were made after removing the remaining resist in a solvent.

### Metalens fabrication.

We use plasma-enhanced chemical vapor deposition (PECVD) to deposit the device layer (1 μm amorphous silicon) on top of MOS cleaned 4-inch fused silica wafer substrate. We then use a Gamma Automatic Coat-Develop tool to spin-coat and pre-bake a 60-nm ARC coating and 600-nm UV210 photoresist on the substrate. With the prefabricated four reticles, we exposed 16 parts, with each part a 20 mm × 20 mm square, by an ASML PAS 5500/300C DUV wafer stepper using the rotation-and-stitching method. After exposing the 16 parts, post-exposure bake and development were followed using the Gamma Automatic Coat-Develop tool to get the resist pattern. After removing the ARC layer with $O_2$ plasma, HBr/Ar gas was used to dry-etch the amorphous silicon in an inductively-coupled-plasma reactive ion etching (ICP-RIE) system to create the nanocylinders. Finally, the photoresist layer and the ARC layer were removed by $O_2$ plasma.

### Optical characterization.

For focusing capabilities measurements, laser output from an optical parametric oscillator (OPO) pumped by a Ti:Sapphire fs-pulse laser system was coupled into a single-mode fiber (Thorlabs P1-SMF28E-FC-10). At the other end of the fiber, an 8-inch-aperture plano-convex lens was used as a beam expander and collimator, which expanded the beam diameter to ~100 mm. The incident beam was much larger than the metalens, which ensures that the entire metalens aperture is illuminated by a plane-wave-like incidence. After the beam passed through the metalens, the focal spot was imaged by a long-working-distance NIR objective lens mounted on a motorized linear translation stage (Newport ILS50CC with ESP 300 controller). Finally, the focal spot images were captured by an InGaAs infrared camera (Sensors Unlimited, GA640CB) placed at the back focal plane of a 2-inch wide-angle tube lens with a 200 mm focal length. The imaging system has a magnification of 50x.

For imaging capability measurements, the USAF resolution test chart was illuminated by a halogen lamp. The light from the lamp was expanded using a 4F system with 2x magnification to get more uniform illumination. The test chart and the IR camera were placed ~ 2 focal lengths away, in front of and behind the metalens, respectively. A bandpass filter (Thorlabs FB1450-12), with a 1450 nm central wavelength and a 12 nm bandwidth, was mounted on the camera to reduce the background noise. The camera's position was fine-tuned by a motorized translation stage (Newport UTS150CC with ESP 300 controller) to ensure the best image quality.

**Author Contributions**

X.N. and M.S. conceived the project. X.N. supervised the study. X.N. and X.C. performed the design and numerical simulations. L.Z. developed the fabrication process and fabricated the large-scale metalenses. S.C. and L.Z. performed metalens characterizations. S.C. developed the stage control code and analysis codes for metalens characterization. Y.Ding and Y.Duan provided technical support for metalens characterization and fabrication. M.R. helped amorphous silicon deposition and SEM imaging of the metalenses. L.Z., S.C., and X.N. wrote the manuscript, and all authors discussed the results and commented on the manuscript.

**Acknowledgments**

The authors acknowledge J. Treichler, G. Bordonaro, M. Skvarla, and J. Clark for their generous help when using the Cornell NanoScale Facility (CNF). This work was performed in part at the Cornell NanoScale Facility, a member of the National Nanotechnology Coordinated Infrastructure (NNCI), which is supported by the National Science Foundation (Grant NNCI-2025233). L.Z. thanks J.-S. Park, M. Song, J. Xiang, and W. Wang for their kind help in scheduling and operating the facilities in CNF. The authors acknowledge M. Labella, G. Lavallee, and B. Liu for their help in using the nanofabrication facility at Penn State. The authors acknowledge partial support from the National Aeronautics and Space Administration Early Career Faculty Award